\documentclass[aps,prb,10pt,twocolumn,letterpaper]{revtex4-1}
\usepackage{amsmath}
\usepackage{dcolumn}
\usepackage{graphicx}
\usepackage{latexsym}
\usepackage{amssymb}
\usepackage{color}

\def\Bp{\mathrel{B_{\phi}} }
\def\Bc{\mathrel{B_{\text{c2}}} }
\def\Bscf{\mathrel{B_{\text{SCF}}} }
\def\tp{\mathrel{\tau_{\phi}} }

\begin{document}

\title{Observation of the ghost critical field for superconducting fluctuations in a disordered TaN thin film}

\author{Nicholas P. Breznay}
\affiliation{Department of Applied Physics, Stanford University, Stanford, CA}
\author{Aharon Kapitulnik} 
\affiliation{Department of Applied Physics, Stanford University, Stanford, CA}
\affiliation{Department of Physics, Stanford University, Stanford, CA}

\date{\today}

\begin{abstract} 
We experimentally study the ghost critical field (GCF), a magnetic field scale for the suppression of superconducting fluctuations, using Hall effect and magnetoresistance measurements on a disordered superconducting thin film near its transition temperature $T_c$. We observe an increase in the Hall effect with a maximum in field that tracks the upper critical field below $T_c$, vanishes near $T_c$, and returns to higher fields above $T_c$. Such a maximum has been observed in studies of the Nernst effect and identified as the GCF. Magnetoresistance measurements near $T_c$ indicate quenching of superconducting fluctuations, agree with established theoretical descriptions, and allow us to extract the GCF and other parameters. Above $T_c$ the Hall peak field is quantitatively distinct from the GCF, and we contrast this finding with ongoing studies of the Nernst effect and superconducting fluctuations in unconventional and thin-film superconductors.
\end{abstract}

\maketitle

\section{Introduction}

The ghost critical field (GCF) is a magnetic field scale for the suppression of fluctuations in a superconductor above its transition temperature $T_c$, analogous to the upper critical field $\Bc$ below $T_c$. The GCF was first observed by Kapitulnik \textit{et al.},~\cite{kapitulnik85pd} who analyzed magnetoconductance measurements of the disordered superconductor InGe and identified a magnetic field $\Bscf$, increasing with temperature above $T_c$, where superconducting fluctuations (SCF) begin to show field dependence. Such a field should correspond to the coherence length $\xi^*$ for SCF, just as the upper critical field $\Bc$ depends on the coherence length $\xi$ below $T_c$ through $\Bc = \frac{\phi_0}{2 \pi \xi^2}$ ($\phi_0$ is the flux quantum). In contrast to $\Bc$, which separates distinct superconducting and normal states, the GCF represents a crossover field between regions of field-independent and field-suppressed SCF. Careful description of the GCF is vital to understand the nature of high-$T_c$ and unconventional superconductors, where there is incomplete understanding of the normal state, fluctuation effects, and possible competing ground states.

\begin{figure}
\centering
\includegraphics[width=1.0\columnwidth]{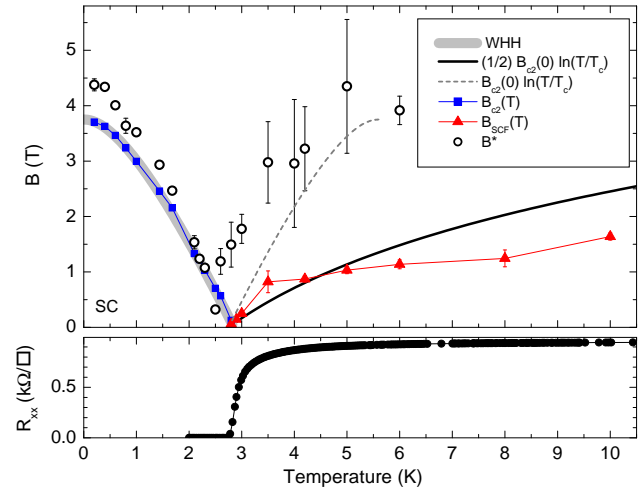}
\caption{ \footnotesize \setlength{\baselineskip}{0.8\baselineskip} (Color online) Upper panel: characteristic magnetic fields versus temperature for a superconducting TaN film, showing the Hall effect peak field $B^*$, the upper critical field $\Bc$, and $\Bscf$, the characteristic field for superconducting fluctuations. The thick gray curve shows the predicted upper critical field below $T_c$, while the continuous curve shows the predicted behavior for the ghost critical field above $T_c$. The broken curve shows another possible estimate for $\Bscf$ as a ``mirror'' of $\Bc$. Lower panel: the resistive superconducting transition over the same temperature range. $R_{xx}$ falls to 0 at $T_c \approx$~2.75~K, where both $\Bscf$ and $\Bc$ also vanish.}
\label{fig:phasediag}
\end{figure}

Since the work of Ref.~\onlinecite{kapitulnik85pd}, the GCF has been tentatively identified and investigated in superconductors using a variety of transport techniques. Many studies using magnetoresistance (MR) measurements have quantitatively identified the GCF in both conventional~\cite{rosenbaum01hc} and high-$T_c$~(Ref.~\onlinecite{rullier11ar}) superconductors. More recently, the Nernst effect has become a common probe of SCF, and has been used to observe amplitude fluctuations over a broad range of temperatures above $T_c$ in NbSi~(Refs.~\onlinecite{pourret06al, pourret07al}) and a crossover between amplitude and phase fluctuations in InO$_{x}$~\cite{spathis08ap}. In addition, Nernst measurements have supported claims of vortex physics in high-$T_c$ compounds above $T_c$~(Ref.~\onlinecite{wang01xk}) and competing order, leading to reductions of $T_c$ and $\Bc$~\cite{chang12dc}. Several of these studies revealed a peak in the Nernst effect in both high-$T_c$~(Ref.~\onlinecite{chang12dc}) and conventional thin film~\cite{pourret06al,spathis08ap} superconductors; this peak was associated with the GCF by many authors, although other interpretations have also been advanced.~\cite{wang06lo} However, no study has observed this feature and made a quantitative comparison with the GCF as determined from the fluctuation magnetoconductance.

In this article, we study the Hall effect in an amorphous superconducting film and observe a temperature dependent peak very similar to that seen in Nernst effect measurements. This Hall effect peak, ocurring at a magnetic field $B^*$, tracks $\Bc$ below $T_c$ and appears comparable to $\Bscf$ above $T_c$. However, quantitative analysis of MR measurements allows us to extract $\Bscf$, and we show that this peak field $B^*$ is distinct from the GCF above $T_c$. Fig.~\ref{fig:phasediag} shows this important result, plotting the temperature dependence of our extracted $B^*$ and $\Bscf$. Our measurements of $\Bscf$ are also consistent with the expected magnitudes of $\xi$ and $\xi^*$ near $T_c$.

In the Ginzburg-Landau (GL) region near $T_c$, $\Bc$ should increase linearly with decreasing temperature with a slope $S = \left.\frac{d \Bc }{dT}\right|_{T_c}$. At $T_c$, $\Bc \rightarrow 0$ and $\xi$ diverges, and above $T_c$, $\xi^*$ decreases with increasing temperature, mirroring $\xi$. $\xi^*$ represents the characteristic size of SCF and is expected to be of order $\xi$. In any mean-field description of a continuous phase transition, the ratio of the susceptibility $\chi$ above and below the transition temperature is universal and equal to 2; since coherence length $\xi \sim \sqrt{\chi}$, we must have $\xi^*/\xi \sim \sqrt{2}$, as shown in Ref.~\onlinecite{tinkham96} in the zero dimensional case. We therefore expect that the slope of critical field for SCF above $T_c$, $S^* = \left.\frac{d \Bscf }{dT}\right|_{T_c}$, to be a factor of 2 smaller than $S$.

We have tested this prediction by studying the GCF in a disordered thin film superconductor. We simultaneously measure the Hall effect and longitudinal resistance as a function of temperature and magnetic field; these data are presented in Sec.~\ref{sec:expt}. In Sec.~\ref{sec:hall} we discuss the peak in Hall effect appearing at a field $B^*$; $B^*$ indicates a crossover from vortex to SCF physics at temperatures below $T_c$ and a field scale for suppression of SCF above $T_c$. We also find a MR consistent with the suppression of SCF as described in Sec.~\ref{sec:mr}. Analysis of magnetoconductance data show excellent agreement with established theory for SCF conductivity corrections and allow extraction of the dephasing field $\Bp$ (and corresponding dephasing time $\tp$) and $\Bscf$. The GCF is consistent with GL theory scaling for $\xi$ above and below $T_c$, saturates at high temperatures $T \gg T_c$, and is distinct from the Hall effect peak field $B^*$ as discussed in Sec.~\ref{sec:disc}.

	\begin{figure}
	\centering
	\includegraphics[width=1.0\columnwidth]{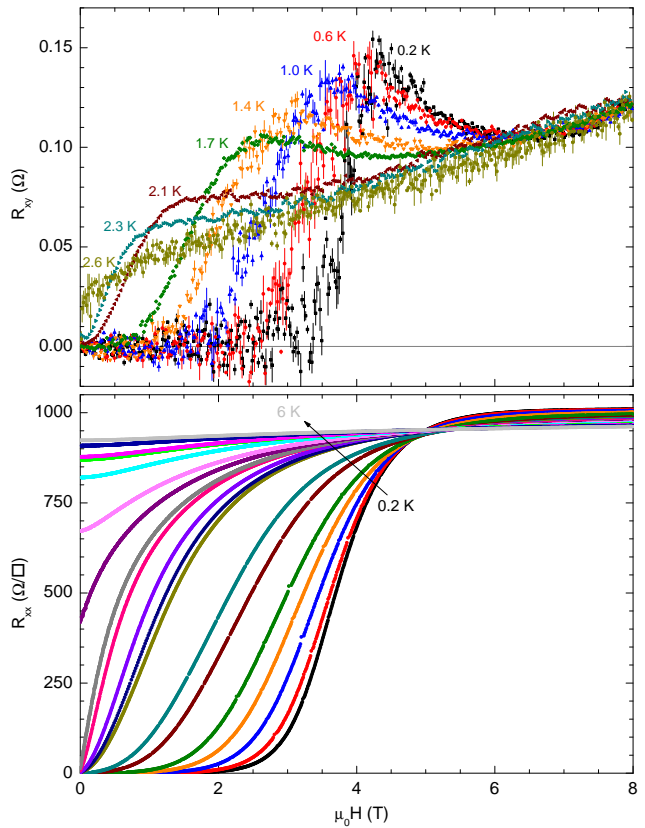}
	\caption{ \footnotesize \setlength{\baselineskip}{0.8\baselineskip} (Color online) The Hall effect $R_{xy}$ (upper panel) and longitudinal resistance $R_{xx}$ (lower panel) versus magnetic field, at temperatures near $T_c \approx$ 2.75 K. $R_{xy}$ shows a maximum near $\Bc$ for temperatures below $T_c$, which moves to zero field as the temperature approaches $T_c$. $R_{xx}$ curves are shown at temperatures of 0.2, 0.4, 0.6, 0.8, 1.0, 1.4, 1.7, 2.1, 2.2, 2.3, 2.5, 2.6, 2.8, 3.0, 3.5, 4.0, 4.2, 5.0, and 6.0 K.}
	\label{fig:hallmr}
	\end{figure}

\section{Experimental results}
\label{sec:expt}

We study a 4.0-nm-thick disordered TaN film described in detail in previous work.~\cite{breznay12mt} Hall measurements indicate a metallic n-type carrier density 8$\times$10$^{22}$ cm$^{-3}$. The normal state sheet resistance at 20 K is $R^n_{xx}$ = 0.94 k$\Omega / \Box$. The lower panel of Fig.~\ref{fig:phasediag} shows the resistive superconducting transition with $T_c \approx$ 2.75 K; the rounding of the transition above $T_c$ is due to enhancement of the conductance from SCF. The upper panel of Fig.~\ref{fig:phasediag} summarizes the main result of this work, showing the magnetic field scales associated with the peak in the Hall effect $B^*$, $\Bc$, and the characteristic field for SCF $\Bscf$, extracted from analysis of the fluctuation magnetoconductance. Each of these quantities, as well as theoretical predictions~\cite{werthamer66hh,tinkham96} for $\Bc$ and $\Bscf$ shown as continuous and broken curves in Fig.~\ref{fig:phasediag}, are explained in the following. Before examining the Hall data and $R_{xx}$ data used to determine $B^*$ and $\Bscf$, let us establish the two-dimensional (2D) nature of this film with respect to superconductivity and localization.

\subsection{Sample dimensionality}

As found previously~\cite{breznay12mt} the zero-temperature coherence length $\xi$(0) is 8.4 nm, and the film can be considered to be 2D with respect to superconductivity and SCF. The dimensionality for localization phenomena such as weak antilocalization (WAL) is determined by the dephasing length $l_{\phi}$, related to the phase breaking time $\tau_{\phi}$ through $L_{\phi} = \sqrt{D \tau_\phi}$. The phase breaking rate is sensitive to both inelastic scattering (with lifetime $\tau_i$) and temperature-independent processes such as spin-flip scattering, however, we find no evidence for this latter contribution to dephasing. Since $\tau_i$ generally increases with decreasing temperature, $\ell_{\phi}$ grows longer as the temperature is reduced and 2D behavior should emerge for sufficiently low $T$. Based on the value of $\ell_{\phi}$~=~10 nm at 20 K determined from WAL magnetoconductance (MC) fits, we conclude that this film is 2D with respect to WAL at and below 20 K.

For sufficiently large magnetic fields the characteristic magnetic length $\ell_B = \sqrt{\hbar/4 e B}$ will be smaller than the film thickness and thus lead to behavior that is not strictly 2D. However, the magnetic length is equal to our $\sim$ 4-nm film thickness at $\sim$ 10 T; this is greater than the maximum magnetic fields used in most of this work ($\sim$ 8 T). Finally, the mean free path $\ell \sim$ 0.1 nm is much less than the film thickness, and so the classical diffusive electronic motion is three dimensional (3D).

\subsection{$R_{xx}$ and $R_{xy}$ measurements}

Figure~\ref{fig:hallmr} shows the measured longitudinal ($R_{xx}$) and Hall ($R_{xy}$) resistances versus applied magnetic field $\mu_0 H$ at temperatures above, near, and below $T_c$. Above $T_c$, $R_{xx}$ is weakly temperature and field dependent; the magnitude and form of the MR are consistent with weak antilocalization (WAL) conductivity corrections, and (approaching $T_c$) additional contributions from SCF. Well below $T_c$ $R_{xx}$ is 0 within the superconducting state, followed by a sharp upturn just below $\Bc$ and saturation in the high field limit. We extract $\Bc$ as the midpoint of these field-tuned resistive transitions, yielding $\Bc$(0) $\sim$ 4 T and a zero-field $T_c$ $\approx$ 2.75 K, comparable to that determined from analysis of the fluctuation conductivity.~\cite{breznay12mt} These values for $\Bc$ are plotted in Fig.~\ref{fig:phasediag} and show good agreement with Werthamer-Helfand-Hohenberg~\cite{werthamer66hh} (WHH) theory for the temperature dependence of $\Bc$ in a disordered superconductor (plotted as a thick gray line on Fig.~\ref{fig:phasediag}). SCF contributions to the diagonal conductivity and magnetoconductance have been studied extensively~\cite{larkinvarlamov2005} and allow for quantitative analysis (described in the following). In contrast, SCF contributions to the Hall effect (and the Nernst effect) remain an active area of theoretical~\cite{michaeli12tf,tikhonov12sf} and experimental~\cite{breznay12mt} study.

For clarity, only $R_{xy}$ data measured at $T \leq T_c$ are shown in Fig.~\ref{fig:hallmr}; these data are discussed in detail in the next section. The corresponding Hall data for $T > T_c$ are presented and analyzed in Ref.~\onlinecite{breznay12mt}; this work demonstrated that an observed enhancement in $R_{xy}$ above $T_c$ can be completely attributed to Gaussian SCF contributions to the Hall conductivity.

\section{Fluctuation Hall effect}
\label{sec:hall}

The normal-state quasiparticle contribution to the Hall effect $R_{xy}^n$ is linear in the applied field as expected in a disordered, single-band metal such as this TaN film. We observe a field-linear $R_{xy}$ at temperatures well above $T_c$ or at the highest fields studied (14 T). Approaching $T_c$ from above, $R_{xy}$ shows a shallow enhancement relative to the normal state at intermediate fields that disappears in the high field limit ($B \gg \Bc$). This enhancement has been shown to arise from SCF contributions to $R_{xy}$~\cite{breznay12mt} and found to be in good agreement with theoretical analysis of Gaussian amplitude fluctuation contributions to the Hall conductivity.~\cite{michaeli12tf,tikhonov12sf} The additional contribution to $R_{xy}$ grows larger in magnitude as $T \rightarrow T_c$ from above, and displays a maximum field $B^*$ that vanishes in this limit. It would be reasonable to expect the Hall peak field $B^*$ to vanish at $T_c$; in this film it appears to vanish at 2.5 K, slightly below $T_c$ but consistent with strong inhomogeniety effects already noted in this film.~\cite{breznay12mt} As the temperature decreases below $T_c$ the Hall effect peak reappears and tracks $\Bc$ to increasing fields. At temperatures below $T_c$ and as the field decreases below $B^*$, the Hall resistance $R_{xy}$ drops quickly to zero.

	\begin{figure}[ht]
	\centering
	\includegraphics[width=1.0\columnwidth]{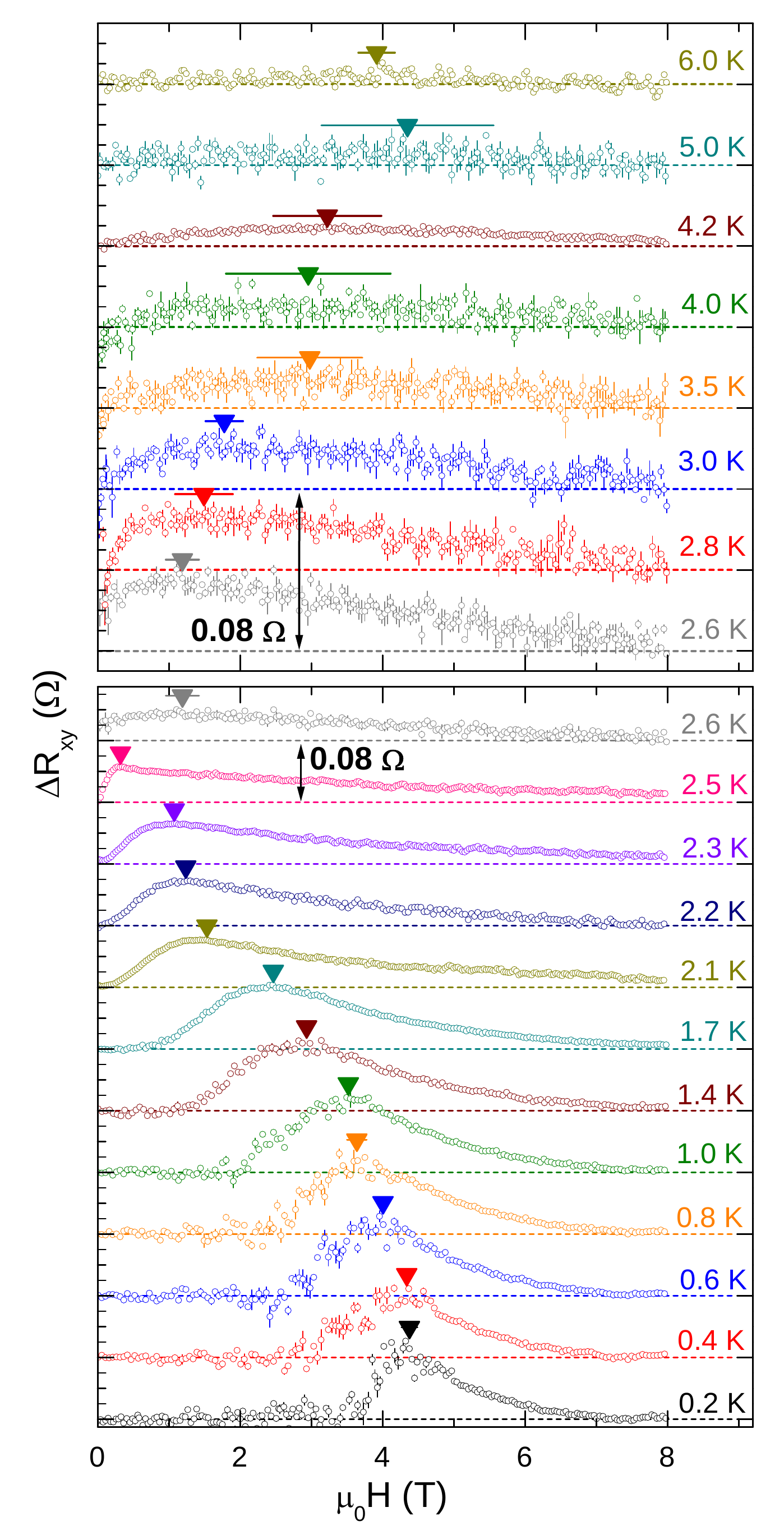}
	\caption{ \footnotesize \setlength{\baselineskip}{0.8\baselineskip} (Color online) Superconducting contribution to the Hall effect $\Delta R_{xy}$ at temperatures above $T_c$ (upper panel) and below $T_c$ (lower panel). The normal state contribution (linear in applied field at $T \gg T_c$) has been subtracted from $R_{xy}$ to obtain $\Delta R_{xy}$. The curves are offset vertically for clarity, and a vertical scale bar of 0.08 $\Omega$ is indicated on both panels. The peak in $\Delta R_{xy}$ occurring at a field $B^*$ (indicated by the triangle above each curve) can be identified through temperatures above and below $T_c$.}
	\label{fig:hallsub}
	\end{figure}

In the superconducting state we expect zero Hall resistance in the presence of particle-hole symmetry and (with a nonzero magnetic field) because vortices are pinned by the strong disorder of this film. As B $\rightarrow \Bc$ from below we expect a nonzero $R_{xy}$ due to vortex motion as well as a crossover to the fluctuation regime at and above $\Bc$. Recent theoretical calculations of SCF contributions to the Hall effect~\cite{michaeli12tf} predict a peak in the Hall resistance near $T_c$ at a field $B_{\text{peak}}$ = 1.3 $\times \Bc$. This is in excellent agreement with the ratio of slope of the Hall peak field near $T_c$ to $S \sim 1.27$.

To further examine the superconducting contributions to the Hall effect $\Delta R_{xy}$, we subtract the normal-state component $R_{xy}^n$ from the $R_{xy}$ data as shown in Fig.~\ref{fig:hallmr}. The contribution to the Hall effect from the normal state is linear in the applied field as verified at high temperature and very high magnetic fields (up to 14 T); see Ref.~\onlinecite{breznay12mt} for further details. Above $T_c$ we subtract this linear component to obtain $\Delta R_{xy}$. Below $T_c$ we estimate $R_{xy}^n = R_H^n B \times \left( \frac{R_{xx}}{R_{xx}^n} \right)^2$. The result $\Delta R_{xy} = R_{xy} - R_{xy}^n$ is shown in Fig.~\ref{fig:hallsub}, for T $< T_c$ (T $> T_c$) in the lower (upper) panel. An enhancement in the Hall effect (indicated by nonzero $\Delta R_{xy}$) appears at high fields for $T > T_c$, decreases to near zero field at $T_c$, and returns to high fields at $T < T_c$.

Figure~\ref{fig:hallsub} also indicates the location of the Hall effect peak $B^*$ for each temperature. We determine $B^*$ using both the highest $\Delta R_{xy}$ value and using the maximum of a local parabolic fit to the data. Uncertainties in $B^*$ reflect the sensitivity of the value to the normal state Hall resistance $R_{xy}^n$ or to the particular fitting algorithm used. $B^*$ tracks $\Bc$ below $T_c$, vanishes near $T_c$, and reappears above $T_c$. These $B^*$ values are also plotted on Fig.~\ref{fig:phasediag}, as mentioned previously. All of the Hall effect isotherms, above and below $T_c$, show a gradual decay of SCF to fields well above $\Bc$(0). The peak field $B^*$ can be followed to temperatures above $2 \times T_c$ and contrasted with Nernst effect measurements on a disordered NbSi thin film that showed traces of SCF up to $30 \times T_c$.~\cite{pourret07al} Having identified the enhancement in $R_{xy}$ due to SCF, we now turn to the magnetoresistance of this film above $T_c$.

\section{Fluctuation magnetoconductance}
\label{sec:mr}

In a superconductor near $T_c$ there may be several contributions to electronic transport arising from normal state quasiparticles, amplitude and phase fluctuations of the superconducting order parameter, and vortex motion. As the temperature is increased from 0 in a superconductor under an applied magnetic field, three distinct regimes of dissipation leading to a nonzero $R_{xx}$ can be identified: unpinned vortex motion at low $T$, phase fluctuations in the vicinity of the Kosterlitz-Thouless transition~\cite{kosterlitz73t}, and finally amplitude fluctuations above the mean-field transition temperature $T_c$. The normal state conductance $G_{xx}$ of a metallic film due to quasiparticles is $G_{xx} \sim \frac{n e^2 \tau}{m^*}$, where $n$ is the sheet carrier density and $m^*$ the effective mass.  Short-lived cooper pairs (SCF) can directly increase the conductivity within their pairing lifetime, as first calculated by Aslamasov and Larkin (AL),~\cite{aslamasov68l} or after they have broken but before they lose phase coherence via the Maki-Thompson (MT) contribution~\cite{maki68,thompson70}. Finally, superconducting vortices also respond to applied field and temperature gradients, giving rise to dissipation and entropy transport. Since a peak in the field-dependent Nernst effect with behavior similar to $B^*$ has been identified as the ghost critical field $\Bscf$,~\cite{pourret07al,spathis08ap,chang12dc} let us consider the fluctuation regime outside of the superconducting state in more detail.

\subsection{Magnetoconductance of a disordered thin film}

The electrical conductance $G_{xx}$ of a disordered metal film in the presence of SCF and WAL corrections can be written
	\begin{equation}
	\label{eq:mconds}
	G_{xx}(B) = G^{n} + \Delta G^{SCF}(B,T) + \Delta G^{WAL}(B,T)
	\end{equation}
where $G^{n}$ is the normal-state Drude conductance, and the remaining two terms arise from superconductivity and disorder-induced localization. (We ignore the classical magnetoconductance $\Delta G/G_0 \sim (\omega_c \tau_{tr})^2$; see the following discussion.) The SCF and WAL corrections in Eq.~\ref{eq:mconds} take on simple forms when the system of interest is 2D and have been studied extensively.~\cite{bergmann84,lee85r,altshuler87ag,larkinvarlamov2005}

The MC due to WAL can be expressed as a function of several characteristic scattering times,~\cite{hikami80ln} and reduces to a simple form for our TaN system. The elastic ($\tau_e$), spin-orbit ($\tau_{so}$), and dephasing ($\tau_{\phi}$) scattering times correspond to characteristic magnetic fields through
	\begin{equation}
	\label{eq:bx}
	B_{x} = \frac{\hbar}{4 e D \tau_x}.
	\end{equation}
With $B_e \sim 10^4$ T as determined from the transport scattering time $\tau_{tr} \sim 10^{-16}$ s, we estimate $B_{so} \sim 10^3$ for interfacial spin-orbit scattering.~\cite{abrikosov62g,meservey78t} Since these characteristic fields are much larger than our experimentally accessible range ($B \sim 10$ T) we consider only the limit $B \ll B_e, B_{so}$. In this case the MC is negative and $\Bp$ is the only free parameter
	\begin{equation}
	\Delta G^{WL}(B) = - \frac{1}{2} \frac{e^2}{2 \pi^2 \hbar} Y\left(\frac{B}{\Bp}\right)
	\label{eq:wlspinorbit}
	\end{equation}
where the function $Y(x)$ is given by
	\begin{equation}
	Y(x) = \ln(x) + \psi\left( \frac{1}{2} + \frac{1}{x} \right)
	\end{equation}
and $\psi$ is the digamma function.

Near a superconducting transition, short-lived SCF provide a parallel conducting channel to the classical Drude conductance and thereby reduce $R_{xx}$ above $T_c$. The Aslamasov-Larkin~\cite{aslamasov68l} (AL) contribution to the conductance arises from direct acceleration of transient Cooper pairs. Quasiparticles from recently decayed Cooper pais can also enhance the conductance while they retain phase coherence; this is the Maki-Thompson~\cite{maki68,thompson70} (MT) contribution. Both of these fluctuation effects are sensitive to applied magnetic fields, leading to two SCF contributions to the MC; these have been studied quantitatively in both conventional~\cite{abraham84r} and unconventional~\cite{ando02a} materials. 

Suppression of the AL conductance channel leads to a negative MC and was calculated by Abrahams, Prange, and Stephen~\cite{abrahams71ps} using time dependent Ginzburg-Landau theory and by Redi~\cite{redi77} from microscopic considerations
	\begin{equation}
	\Delta G^{AL}(B) = 	\frac{e^2}{16\hbar \epsilon} \left( 8 z^2 A(z) - 1 \right).
	\label{eq:almagneto}
	\end{equation}
Here the reduced temperature $\epsilon \equiv \ln\left(\frac{T}{T_c}\right)$, the function $A(z)$ is given by
	\begin{equation}
	A(z) = \psi\left( \frac{1}{2}  + z \right) - \psi(1 + z) + \frac{1}{2z},
	\end{equation}
z $\equiv \Bscf/B$, and $\Bscf$ is the characteristic field associated with SCF. This definiton of $\Bscf$ can by obtained by inserting the Ginzburg-Landau time $\tau_{GL} = \frac{\pi \hbar}{8 k_B T \ln(T / T_c}$ into Eq.~\ref{eq:bx} above.

The AL contribution to the conductivity is dominant close to $T_c$ ($\epsilon \ll 1$), while further above $T_c$ the MT channel can be significant. Larkin~\cite{larkin80} originally calculated this MT term; Lopes dos Santos and Abrahams~\cite{lopes85a} expanded the result:
	\begin{widetext}
	\begin{equation}
	\Delta G^{MT-LdS}(B) = - \beta_{LdS} \frac{e^2}{2\pi^2 \hbar} \left( \psi\left( \frac{1}{2}
	+ \frac{\Bp}{B} \right) - \psi\left( \frac{1}{2} + \frac{\Bscf}{B} \right) + \ln\frac{\Bscf}{\Bp} \right)
	\label{eq:mtlds}
	\end{equation}
	\end{widetext}
where the parameter $\beta_{LdS} \equiv \pi^2 / 4 (\epsilon - \delta)$. The quantity $\delta$ is a cutoff parameter given by
	\begin{equation}
	\delta = \frac{\pi \hbar}{8 k_B T}\frac{1}{\tau_{\phi}},
	\end{equation}
thus $\delta$ is specified by $\Bp$ and known parameters. The combined effects of WAL, AL fluctuations, and MT fluctuations are included within Eqs.~\ref{eq:almagneto} and ~\ref{eq:mtlds}. The range of magnetic fields and temperatures where these expressions are valid is limited; the most restrictive condition (from the MT MC) is~\cite{lopes85a}
	\begin{equation}
	4 D e B \ll k_B T \ln\left( T/T_c \right)
	\label{eq:valid}
	\end{equation}
at temperatures well above $T_c$; near $T_c$ the range of validity increases to $4 D e B \ll k_B T$. We consider all of these theoretical results when analyzing the magnetoresistance data presented next.

	\begin{figure}
	\centering
	\includegraphics[width=0.95\columnwidth]{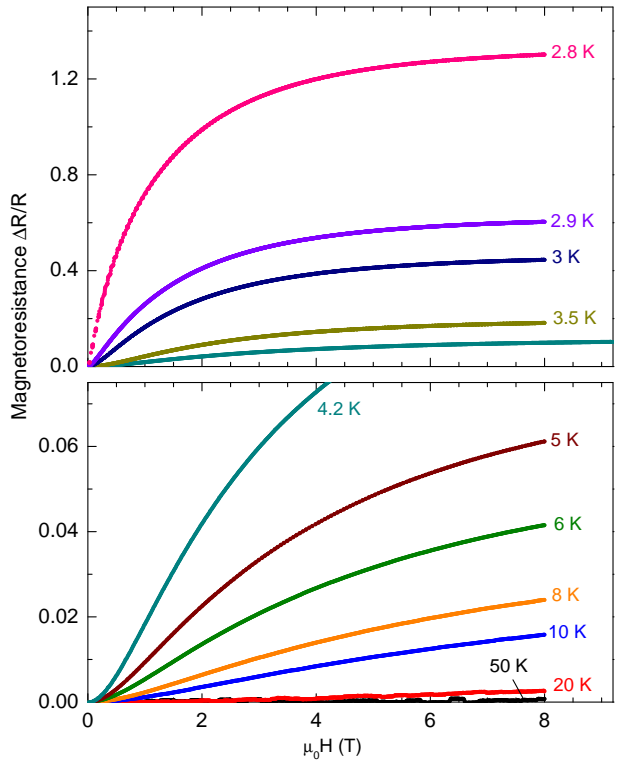}
	\caption{ \footnotesize \setlength{\baselineskip}{0.8\baselineskip} (Color online) Magnetoresistance $\Delta R/R \equiv (R_{xx}(B) - R_{xx}(0))/R_{xx}(0)$ versus applied magnetic field for a superconducting TaN film at temperatures above $T_c$. The lower panel shows temperatures $T \gg T_c$ (50, 20, 10, 8, 6, 5, 4.2 K), and the upper panel shows $T \sim T_c$ (4.2, 3.5, 3.0, 2.9, 2.8 K) with an enlarged vertical scale.}
	\label{fig:mrsupp}
	\end{figure}

\subsection{Magnetoresistance data and fitting results}

Figure~\ref{fig:mrsupp} shows the measured magnetoresistance $\Delta R / R \equiv (R(B)-R(0))/R(0)$ versus applied magnetic field at temperatures well above $T_c$ (lower panel) and near $T_c$ (upper panel). At sufficiently high temperatures the MR should scale using Kohler's rule for this disordered film; the transport scattering time  $\tau_{tr} \sim 10^{-16}$ s and thus the magnetoresistance should be $\sim (\omega_c \tau)^2 \approx 10^{-8}$ in an applied field of 8 T. The MR at 50 K is negligible, consistent with this prediction. At lower temperatures, the MR becomes nonzero; at 20 K it is positive and the magnitude and shape are consistent with WAL. Approaching $T_c$ (below 20 K) the MR remains positive and increases in size due to field suppression of SCF; nearing $T_c$ the MR saturates at high field and becomes of order 1.

	\begin{figure}
	\centering
	\includegraphics[width=1.0\columnwidth]{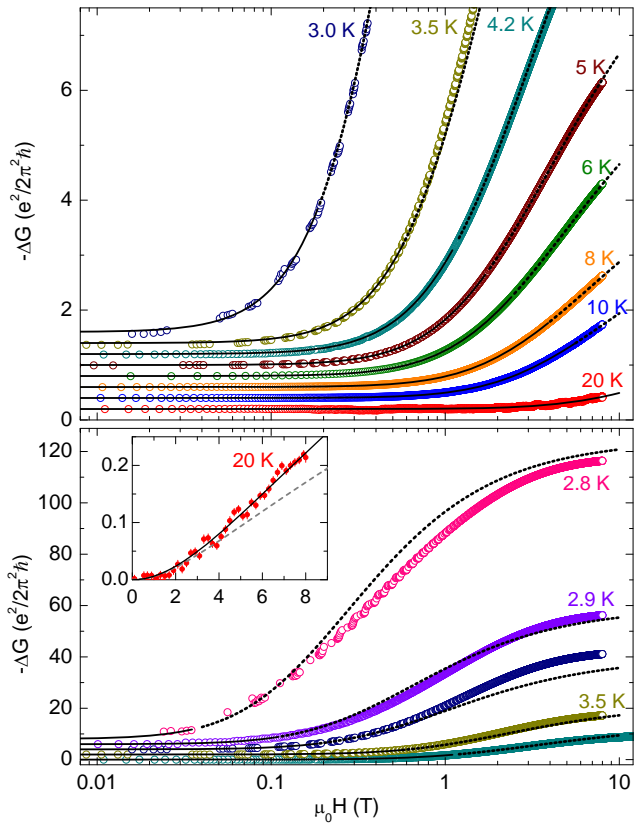}
	\caption{ \footnotesize \setlength{\baselineskip}{0.8\baselineskip} (Color online) Negative magnetoconductance $-\Delta G$ versus magnetic field plotted on a log scale at temperatures above $T_c$ (upper panel) and near $T_c$ (lower panel), as well as fits (continuous and dotted curves) to theories for the magnetic field suppression of superconducting fluctuation and weak antilocalization contributions to the conductance (see text). The lower panel shows 4.2, 3.5, 3.0, 2.9, and 2.8 K data, and the upper panel shows 20, 10, 8, 6, 5, 4.2, and 3.5 K data. The inset shows the 20 K data and full fit (continuous curve), and also a fit to only WAL theory (broken curve).}
	\label{fig:gcfMR}
	\end{figure}

To quantitatively analyze this low-temperature behavior we consider the magnetoconductance (MC). Since the Hall angle is negligible, we calculate the negative magnetoconductance $-\Delta G(B)$
	\begin{equation}
	-\Delta G(B) = G(0) - G(B) = \frac{R_{xx}(B) - R_{xx}(0)}{R_{xx}(B)R_{xx}(0)}.
	\label{eq:nmc}
	\end{equation}
Figure~\ref{fig:gcfMR} shows the negative MC as a function of magnetic field (on a log scale) for temperatures near (lower panel) and well above (uppper panel) $T_c$; the curves are offset for clarity and the upper and lower panels show different vertical scales. At 20 K where we expect no traces of superconductivity (bottom curve of upper panel, and inset) the MC is positive and consistent with WAL. (We expect weak antilocalization in a film with large average atomic number such as TaN.~\cite{geier92b}) The inset of Fig.~\ref{fig:gcfMR} shows a fit to the combined WAL and MT MC (Eq.~\ref{eq:wlspinorbit} and Eq.~\ref{eq:mtlds}) as the solid curve, yielding a dephasing field $\Bp$ = 2.5 T. (Such a MT contribution to the MC well above $T_c$ has been observed previously; see e.g. Ref.~\onlinecite{rosenbaum85}.) We also show a fit to just the WAL contribution (Eq.~\ref{eq:wlspinorbit}) as the broken curve; the extracted dephasing field $\Bp$ = 1.5 T in this case. Taking D = 0.5 cm$^2$/s, we estimate $\tp$ = 2.2 $\times 10^{-12}$ s using $\Bp = \frac{\hbar}{4 e D \tp}$; this value is comparable to the predicted time for electron-electron dominated inelastic scattering ~\cite{altshuler82ak} of $\tp = 4 \times 10^{-12}$. The corresponding dephasing length $\ell_{\phi}$ is 10 nm, larger than the film thickness and confirming that WAL effects are 2D.

At temperatures below 20 K, the MC data show additional contributions due to SCF. We fit these traces to Equations~\ref{eq:almagneto}, ~\ref{eq:mtlds} and ~\ref{eq:wlspinorbit}, which include both SCF and WAL effects and require only two free parameters: $\Bp$ and $\Bscf$. (We fix $T_c = 2.75$ K, and use the explicit definitions for $\beta_{LdS}$ and $\delta$ given above.)

	\begin{figure}[t]
	\centering
	\includegraphics[width=1\columnwidth]{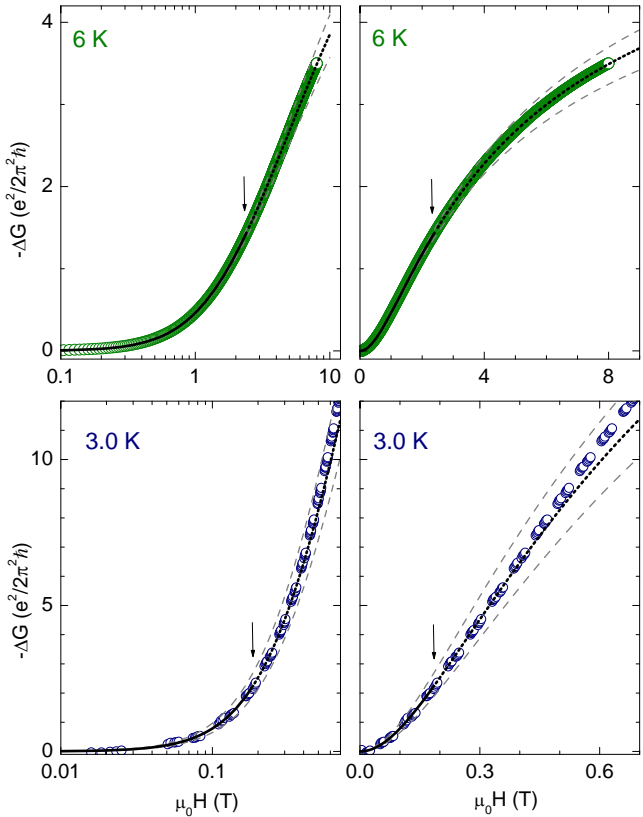}
	\caption{ \footnotesize \setlength{\baselineskip}{0.8\baselineskip} (Color online) Negative magnetoconductance data, along with fits to the combined theories of weak antilocalization and superconducting fluctuations. The upper panels show 6-K results on logarithmic (left) and linear (right) magnetic field scales; the lower panels show similar plots for 3-K data. The continuous and dotted curves show the best fit; the dashed curves use the best-fit value of $\Bp$ but vary $\Bscf$ by $\pm 10 \%$ from the best-fit value. Arrows indicate the limit of validity of the Aslamsov-Larkin magnetoconductance theory (see text).}
	\label{fig:twofits}
	\end{figure}

\begin{figure}[b]
	\centering
	\includegraphics[width=0.8\columnwidth]{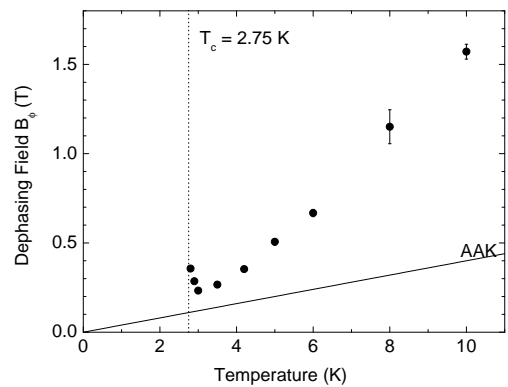}
	\caption{ \footnotesize \setlength{\baselineskip}{0.8\baselineskip} Dephasing field $\Bp$ extracted from our analysis of low-temperature magnetoconductance measurements. The dotted line indicates $T_c$, and the solid curve is an estimate for T-linear dephasing due to electron-electron scattering from the theory of Altshuler, Aronov, and Khmelnitski (AAK).~\cite{altshuler82ak}}
	\label{fig:bphi}
	\end{figure}

The resulting best fits to the combined magnetoconductance expressions are plotted as the solid and dotted curves in Fig.~\ref{fig:gcfMR}, and are in excellent agreement with the MC above $T_c$ over a wide range of temperatures and magnetic fields. We restrict the fitting range to the region of validity give by Eq.~\ref{eq:valid}, and vary this range by up to a factor of 4 to obtain an estimate of the uncertainty in the fitting parameters. Note that the best-fit curves are plotted as solid curves up to their limit of validity, and dotted curves above this. We determine both $\Bscf$ (plotted in Fig.~\ref{fig:phasediag}) and $\Bp$ (shown in Fig.~\ref{fig:bphi}) as a function of temperature; these parameters are discussed further below.

To further examine the fit quality and illustrate the effect of variation in the parameters, Fig.~\ref{fig:twofits} shows MC data, the best fit results (continuous and dotted black curves) on both logarithmic (left) and linear (right) field scales for two temperatures, 6 K (upper panels) and 3 K (lower panels). The upper limit for the validity of the MT fitting expression is also indicated by the arrow. These best-fit curves show excellent agreement at low fields (continuous curves) for all data sets, and only modest deviations at high fields well above the range of validity of the theoretical expressions. Since we are particularly interested in the parameter $\Bscf$, we also show the effect of increasing or decreasing $\Bscf$ by 10\% while keeping $\Bp$ constant; these are shown as the broken curves in each panel, and indicate the sensitivity of the fitting procedure to $\Bscf$.

Figure~\ref{fig:bphi} shows the best fit values of $\Bp$ as a function of temperature. Far above $T_c$ $\Bp$ increases with temperature, as expected for dephasing arising from $T$-dependent electron-electron and electron-phonon scattering processes.~\cite{lin02b} For a disordered conducting film, at sufficiently low $T$, the dephasing should be dominated by electron-electron scattering with a scattering rate proportional to $R_{xx}$ and given by~\cite{altshuler82ak}
	\begin{equation}
	\tau_{i,(e-e)}^{-1} = \frac{e^2}{2 \pi^2 \hbar} R_{xx} \frac{\pi k_B T}{\hbar} \ln{\left( \frac{\pi \hbar}{e^2 R_{xx}} \right)}.
	\end{equation}
Using this expression we estimate the electron-electron scattering contribution to $\Bp$ at low temperatures; this estimate is plotted as the solid curve in Fig.~\ref{fig:bphi}. With increasing temperatures we expect electron-phonon scattering to become the dominant dephasing process, leading to a stronger than $T$-linear temperature dependence. Such an increase is evident at the highest temperatures in Fig.~\ref{fig:bphi}. $\Bp$ shows a small upturn very close to $T_c$; this result (although sensitive to the analysis procedure used) is consistent with enhanced dephasing due to electron-SCF scattering~\cite{brenig86ph} and will be examined carefully in future work.~\cite{brezkap}

\section{Discussion}
\label{sec:disc}

Having extracted $\Bc$ and $\Bscf$, we consider the behavior of $\xi$ in further detail. $\Bc$ and $\Bscf$ mirror one another above and below $T_c$. Below $T_c$, $\Bc$ and the corresponding $\xi$ follow the WHH~\cite{werthamer66hh} behavior (the thick gray curve in Fig. 1). Above $T_c$, we expect $\xi^*$ to be a factor of $\sqrt{2}$ larger than below $T_c$~\cite{tinkham96} and therefore $\Bscf$ to be a factor of 2 smaller than $\Bc$ (solid curve in Fig. 1). Both $\Bc$ and $\Bscf$ are linear in the vicinity of $T_c$, and their linear slopes (1.8 and 1.0) near $T_c$ are comparable to this factor of 2. While $\Bscf$ and $\Bc$ vanish at 2.75 K, this is somewhat above the temperature $T^*$ where the peak in the Hall effect decreases to zero field. However, we have already seen indications of strong inhomogeneity effects in this film~\cite{breznay12mt}, and while $R_{xx}$ may decrease to zero at the point where a percolation path emerges, the Hall effect should be sensitive to global superconductivity occurring through the film. (We also expect a 2D film with $R_{xx} \sim$ 1 $k\Omega$  to show a Kosterlitz-Thouless-Berezinskii (KTB) vortex unbinding temperature $T_{KTB}$ that is $\sim$ 0.1 K below $T_c$~\cite{beasley79mo}, but this is less than the observed separation between $T_c$ and $T^*$.)

Recent studies of the Nernst effect in low-$T_c$~\cite{pourret07al, spathis08ap} and high-$T_c$~\cite{chang12dc} superconductors have associated a peak in the Nernst signal with the GCF. While the Nernst effect may be sensitive to both amplitude and phase fluctuation effects, it is argued to be sensitive to the presence of vortex-like excitations in the fluctuation regime~\cite{wang01xk} or vortex motion within the superconducting state~\cite{vidal73}. In particular, several Nernst studies observe a peak field $B^*$ comparable to or larger than $\Bc$ in the vicinity of $T_c$. This is consistent with theoretical predictions for the Hall effect~\cite{michaeli12tf} and with our findings, however it is inconsistent with the GL theory scaling for $\xi$ above and below $T_c$ that should relate $\Bc$ and the GCF.~\cite{tinkham96}  Consequently, we speculate that the crossover field observed in the Nernst studies may similarly represent a different characteristic field scale than the GCF as first described by Kapitulnik \textit{et al.}~\cite{kapitulnik85pd}

To summarize, we have studied the magnetic field scales for the suppression of SCF in a disordered thin film. We observe an enhancement in the Hall effect at temperatures above and below $T_c$, similar in behavior to that seen in several recent studies of transport in disordered and high-$T_c$ superconductors. Simultaneous measurement of the longitudinal MR, and quantitative analysis using theories for WAL and SCF contributions to the MC, allow robust extraction of both the dephasing field and the characteristic SCF suppression field $\Bscf$. We find that $\Bc$ is distinct from the peak field the appears in the Hall effect and in good agreement with expected behavior in the vicinity of $T_c$.

We thank K. Michaeli, A. Palevski, and F. Lalibert\'e for helpful conversations, E. Schemm for comments on the manuscript, and M. Tendulkar for sample preparation and characterization assistance. This work supported by the National Science Foundation grant NSF-DMR-9508419.

\end{document}